\documentstyle[a4wide,11pt]{article}

\begin{document}

\pagestyle{plain}

\newcommand{\be}{\begin{equation}}
\newcommand{\ee}{\end{equation}}
\newcommand{\bear}{\begin{eqnarray}}
\newcommand{\ear}{\end{eqnarray}}
\newcommand{\no}{\noindent}

\date{}

\renewcommand{\theequation}{\arabic{section}.\arabic{equation}}
\newcommand{\half}{{1\over 2}}
\newcommand{\Dhalf}{{D\over 2}}
\newcommand{\e}{\mbox{e}}
\newcommand{\g}{\mbox{g}}
\renewcommand{\arraystretch}{2.5}
\newcommand{\GeV}{\mbox{GeV}}
\newcommand{\cL}{\cal L}
\newcommand{\D}{\cal D}
\newcommand{{\Tr}}{\rm Tr}
\newcommand{{\tr}}{\rm tr}
\newcommand{\Det}{\rm Det}
\newcommand{\PP}{\cal P}
\newcommand{\G}{{\cal G}}
\newcommand{\Gz}{{G^z_{Bab}}}
\newcommand{\Gzp}{{\dot G^z_{Bab}}}

\def\R{1\!\!{\rm R}}
\def\eins{  1\!{\rm l}  }

\newcommand{\symb}{\mbox{symb}}
\renewcommand{\arraystretch}{2.5}
\newcommand{\slD}{\raise.15ex\hbox{$/$}\kern-.57em\hbox{$D$}}

\pagestyle{empty}

\renewcommand{\thefootnote}{\fnsymbol{footnote}}

\hfill {\sl HUB-EP-97/25}
\vskip-.1pt
\hfill {\sl HD-THEP-97/14}

\vskip 2.4cm
\begin{center}
{\bf\Large The Two-Loop Euler-Heisenberg Lagrangian}\medskip\\
{\bf\Large in Dimensional Renormalization}
\vskip1cm

\vskip.7cm
{\large D. Fliegner$^{\rm \,\,a}$
        \footnote{e-mail address D.Fliegner\@ThPhys.Uni-Heidelberg.De},
        M.Reuter$^{\rm \,\,b}$, 
        M. G. Schmidt$^{\rm \,\,a}$
        \footnote{e-mail address M.G.Schmidt\@ThPhys.Uni-Heidelberg.De}, 
        C. Schubert$^{\rm \,\,c}$
        \footnote{e-mail address schubert@qft2.physik.hu-berlin.de} 
        \footnote{Supported by Deutsche Forschungsgemeinschaft} }

\vskip.7cm 

{\it $^{\rm a}$Institut f\"ur Theoretische Physik,
Universit\"at Heidelberg\\
Philosophenweg 16,
D-69120 Heidelberg, Germany}

\vskip.5cm

{\it $^{\rm b}$Deutsches Elektronen-Synchrotron DESY,\\
Notkestrasse 85, D-22603 Hamburg, Germany}

\vskip.5cm

{\it $^{\rm c}$Institut f\"ur Physik,
Humboldt Universit\"at zu Berlin\\
Invalidenstr. 110, D-10115 Berlin, Germany}

\vskip1.5cm

{\large\bf Abstract}

\end{center}

\begin{quotation}

\noindent
We clarify a discrepancy between two previous calculations
of the two-loop QED Euler-Heisenberg Lagrangian, both
performed in proper-time regularization, by calculating this
quantity in dimensional regularization.

\end{quotation}

\clearpage

\renewcommand{\thefootnote}{\protect\arabic{footnote}}
\pagestyle{plain}
\renewcommand{\theequation}{\arabic{equation}}
\setcounter{equation}{0}
\setcounter{page}{1}
\setcounter{footnote}{0}

One of the earliest results in quantum
electrodynamics was Euler and Heisenberg's 
~\cite{eulhei} calculation
of the one-loop effective Lagrangian induced
by an electron loop for an electromagnetic
background field with constant field
strength tensor $F_{\mu\nu}$. Written in
Schwinger's proper-time representation 
\cite{schwinger51}, this Lagrangian reads

\begin{equation}
{\cal L}_{\rm spin}^{(1)}[F]=
-{1\over 8\pi^2}\int_0^{\infty}
{dT\over T^3}
{\rm e}^{-m^2T}
{(eaT)(ebT)
\over\tan(eaT)\tanh(ebT)}
\, .\\
\label{eulheispin}\nonumber
\end{equation}
\no
Here $a$ and $b$ may be expressed in
terms of the two invariants of
the electromagnetic field,

\begin{eqnarray}
a^2&=&\frac{1}{2}
\biggl[
{\bf E}^2-{\bf B}^2
+\sqrt
{{({\bf E}^2-{\bf B}^2)}^2
+4{({\bf E}\cdot
{\bf B})}^2
}
\biggr]\nonumber\\
b^2&=&\frac{1}{2}
\biggl[
-({\bf E}^2-{\bf B}^2)
+\sqrt
{{({\bf E}^2-{\bf B}^2)}^2
+4{({\bf E}\cdot
{\bf B})}^2
}
\biggr]
\, .
\label{defab}
\end{eqnarray}
\noindent
Schwinger also supplied the corresponding result for
scalar quantum electrodynamics ~\cite{schwinger51}, 

\begin{equation}
{\cal L}_{\rm scal}^{(1)}[F]=
{1\over 16\pi^2}\int_0^{\infty}
{dT\over T^3}
{\rm e}^{-m^2T}
{(eaT)(ebT)
\over\sin(eaT)\sinh(ebT)}
\, . \\
\label{eulheiscal}\nonumber
\end{equation}
\no
Even the first radiative correction to the
Euler-Heisenberg Lagrangian, due to the exchange
of one internal photon in the electron loop,
has been obtained many years ago 
by Ritus ~\cite{ritusspin} in terms of
a two-parameter integral.

Dittrich and one of the
authors ~\cite{ditreu} later
obtained a similar but simpler
integral representation for the same quantity.
They also verified the agreement of both
representations in the strong field limit.

In a recent publication ~\cite{rescsc}
three of the present authors 
showed that this
type of calculation can be
considerably simplified
using the
worldline path integral variant
~\cite{strassler,ss1,ss2,ss3,dashsu,rolsat,zako} of the
Bern-Kosower formalism ~\cite{bk,berdun}. 

This third calculation led to exactly the same
parameter integral 
for the {\sl regularized} effective Lagrangian
as the Dittrich-Reuter calculation, however in
a more elegant way. 

All three calculations were performed in four
dimensions, proper-time regularization, and
on-shell renormalization.
This choice of regularization keeps the integrations
simple, but at the two-loop level
makes it already somewhat nontrivial to
achieve a consistent on-shell renormalization.
The difficulty is due to the non-universal
nature of the proper-time cutoff, and 
somewhat similar to
the problems encountered in multiloop 
Feynman diagrams calculations
performed with a naive momentum space cutoff
(see \cite{vladimirov} and references therein).

In fact, the calculation in ~\cite{rescsc} was
incomplete in so far as we were not able there
to determine the finite part of the
one-loop mass displacement appropriate
to the present calculational scheme.
Moreover, as a byproduct of this investigation
we found that the formulas obtained by Ritus and
Dittrich-Reuter for the renormalized
effective Lagrangians are incompatible as they stand,
and for the very same reason;
if at all,
they can be identified only after a certain finite mass
renormalization. One had to conclude that in
(at least) one of the two previous calculations the 
physical renormalized electron mass had been misidentified
(the strong field limit 
checked in ~\cite{ditreu} is not sensitive to this
discrepancy).

In the present letter, we clarify this matter by
recalculating this effective Lagrangian
 in dimensional regularization.
This paper should thus be seen as supplementing ~\cite{rescsc},
and the reader is referred to that
publication for some of the details of the
formalism used here.

As usual in applications of the ``string-inspired'' technique
to QED \cite{zako}, 
our calculation of the two-loop Euler-Heisenberg
Lagrangian for spinor QED will yield the corresponding
Lagrangian for scalar QED as a partial result. We 
therefore consider the scalar QED case first.

In scalar QED, our starting point is the following
worldline path integral representation for the
bare dimensionally regularized
two-loop Euler-Heisenberg Lagrangian ~\cite{rescsc,ss3,dashsu}, 

\begin{eqnarray}
{\cal L}^{(2)}_{\rm scal}
\lbrack A\rbrack &  = &
-{e^2\over 2}
{{\Gamma (\Dhalf - 1 )}\over {4{\pi}^{D\over 2}}}
{\displaystyle\int_0^{\infty}}
{dT\over T}
e^{-m^2T}
\int_0^T\, d\tau_a 
\int_0^T\, d\tau_b
\nonumber\\
&\times & \hspace{-.6cm}
{\displaystyle\int\limits_{\hspace{.5cm}y(T)=y(0)}} 
\hspace{-.6cm} {\cal D} y \,
{\dot y(\tau_a)
\cdot
\dot y(\tau_b)
\over
{\Bigl({\bigl[y(\tau_a) - y(\tau_b)\bigr]}^2\Bigr)}^
{{D\over 2}-1}}
\,{\rm exp}\biggl [- \int_0^T d\tau
\Bigl ({1\over 4}{\dot y}^2 
+ ie\,
\dot y^{\mu}A_{\mu}(x_0+y)
\Bigr )
\biggr ]
.
\label{2loopscalarpi}
\end{eqnarray}
Here $T$ denotes the usual Schwinger proper-time
parameter for the scalar circulating in the loop.
The path integral ${\cal D}y$ is to be
performed over the space of all closed loops in
$D$ - dimensional spacetime,
with an arbitrary but fixed center of mass
$x_0$,
and traversed in the fixed
proper-time $T$. 
The parameters $\tau_{a,b}$ 
parametrize the end-points
of the photon exchanged in the scalar loop.
This form of the photon insertion corresponds 
to Feynman gauge.
We use euclidean conventions both on the worldline
and in spacetime.

This representation 
of the effective action in terms of a first-quantized
path integral
goes essentially
back to Feynman ~\cite{feynman}, except that we have
dimensionally continued it to $D$ dimensions.

It is useful to take the background field
$A$ in Fock-Schwinger gauge centered at the
loop center of mass $x_0$ ~\cite{ss1}, 
where one has
$A_{\mu}(x_0+y)=\half y^{\nu}F_{\nu\mu}$.
Exponentiating 
the denominator of the photon insertion
using a proper-time
parameter $\bar T$, one arrives at 

\begin{eqnarray}
{\cal L}^{(2)}_{\rm scal}
\lbrack F\rbrack &  = &
-{e^2\over 2}
{\displaystyle\int_0^{\infty}}
{dT\over T}
e^{-m^2T}
\int_0^{\infty}\, d\bar T
{(4\pi\bar T)}^{-{D\over 2}}
\int_0^T\, d\tau_a 
\int_0^T\, d\tau_b
\nonumber\\
&\times&
{\displaystyle\int}
{\cal D} y
\;
\dot y_a
\cdot
\dot y_b
\,
{\rm exp}\biggl [- \int_0^T d\tau
\Bigl ({1\over 4}{\dot y}^2 
+ {i\over 2}e\,
y^{\mu}F_{\mu\nu}\dot y^{\nu}
\Bigr )
-
{{(y_a-y_b)}^2
\over
4\bar T}
\biggr ]
.
\label{2loopscalarpiexpon}
\end{eqnarray}
\no
The new path integral is gaussian,
so that its evaluation amounts
to a single Wick contraction of 
$\langle \dot y_a\cdot \dot y_b\rangle$.
This leads to \cite{rescsc}

\begin{eqnarray}
{\cal L}^{(2)}_{\rm scal}
[F]&=&-
{(4\pi )}^{-D}
{e^2\over 2}
\int_0^{\infty}{dT\over T}e^{-m^2T}T^{-{D\over 2}} 
\int_0^{\infty}d\bar T 
\int_0^T d\tau_a
\int_0^T d\tau_b
\nonumber\\
&\phantom{=}&\times
{\rm det}^{-{1\over 2}}
\biggl[{\sin(eFT)\over {eFT}}
\biggr]
{\rm det}^{-{1\over 2}}
\biggl[
\bar T 
-{1\over 2}
{\cal C}_{ab}
\biggr]
\langle
\dot y_a\cdot\dot y_b\rangle
\quad .
\label{Gamma2scal}
\end{eqnarray}
\noindent
Of the (Lorentz) 
determinant factors appearing here,
the first one
yields just the one-loop Euler-Heisenberg-Schwinger
integrand eq.~(\ref{eulheiscal}).
It represents the dependence of the path integral
determinant on the external field
~\cite{ss1}, while
the second determinant factor
represents its dependence on the photon
insertion.
The Wick contraction is given by
\cite{rescsc}

\begin{equation}
\langle
\dot y_a\cdot\dot y_b\rangle
=
{\rm tr}
\biggl[
\ddot{\cal G}_{Bab}+{1\over 2}
{(\dot {\cal G}_{Baa}-\dot {\cal G}_{Bab})
(\dot {\cal G}_{Bab}-\dot {\cal G}_{Bbb})
\over
{\bar T -{1\over 2}{\cal C}_{ab}}}
\biggr]\; ,
\label{wickscal}
\end{equation}
\noindent
where

\be
{\cal G}_{B}(\tau_1,\tau_2) =
{1\over 2{(eF)}^2}\biggl({eF\over{{\rm sin}(eFT)}}
{\rm e}^{-ieFT\dot G_{B12}}
+ieF\dot G_{B12} -{1\over T}\biggr)
\label{defGF}
\ee
\no
is the bosonic one-loop worldline Green's function
modified by the constant field 
~\cite{rescsc,adlsch,cdd,shaisultanov},
and
$
{\cal C}_{ab}\equiv
{\cal G}_{Baa}
-{\cal G}_{Bab}
-{\cal G}_{Bba}
+{\cal G}_{Bbb}
$.
${\cal G}_B$
generalizes the ordinary worldline Green's
function $G_B$,

\be
G_B(\tau_i,\tau_j)=\mid \tau_i-\tau_j\mid 
-{{(\tau_i-\tau_j)}^2\over T}
\, . 
\label{defG}
\ee
\no
We will often abbreviate $G_{B12}\equiv
G_B(\tau_1,\tau_2)$ etc., and a
``dot'' always denotes a derivative with respect to
the first variable, 
e.g. $\dot G_{B12}={\rm sign}(\tau_1 - \tau_2)
- 2 {{(\tau_1 - \tau_2)}\over T}$.

Performing a partial integration with respect to $\tau_a$ 
on the first term in eq.~(\ref{wickscal})
one can derive the alternative parameter integral

\begin{eqnarray}
{\cal L}^{(2)}_{\rm scal}[F]&=&
-
{(4\pi )}^{-D}
{e^2\over 2}
\int_0^{\infty}{dT\over T}e^{-m^2T}T^{-{D\over 2}} 
\int_0^{\infty}d\bar T 
\int_0^T d\tau_a
\int_0^T d\tau_b
\nonumber\\
&\phantom{=}&\times
{\rm det}^{-{1\over 2}}
\biggl\lbrack{\sin(eFT)\over {eFT}}
\biggr\rbrack
{\rm det}^{-{1\over 2}}
\biggl[
\bar T -{1\over 2}
{\cal C}_{ab}
\biggr\rbrack\nonumber\\
&\phantom{=}&\times
{1\over 2}
\Biggl\lbrace
{\rm tr}\dot{\cal G}_{Bab}
{\rm tr}
\biggl\lbrack
{\dot{\cal G}_{Bab}
\over
{\bar T -{1\over 2}{\cal C}_{ab}}}
\biggr\rbrack
+{\rm tr}
\biggl[
{(\dot {\cal G}_{Baa}-\dot {\cal G}_{Bab})
(\dot {\cal G}_{Bab}-\dot {\cal G}_{Bbb})
\over
{\bar T -{1\over 2}{\cal C}_{ab}}}
\biggr]
\Biggr\rbrace\; .
\label{Gamma2scalpI}
\end{eqnarray}
\noindent
For our present purpose, we can restrict ourselves to the
pure magnetic field case.
This field we take along the z-axis, so that
$F^{12}=B$, $F^{21}=-B$ are the only
non-vanishing components of
the dimensionally continued field strength tensor. 

\no
We also introduce the following abbreviations,

\bear
z&\equiv& eBT\nonumber\\
\gamma&\equiv&{(\bar T +G_{Bab})}^{-1}\nonumber\\
\gamma^z&\equiv&{(\bar T +G_{Bab}^z)}^{-1}\nonumber\\
\label{defzgamma}\nonumber
\end{eqnarray}
\noindent
and the $z$ -- dependent Green's function 
$G_{Bab}^z$,

\bear
G_{Bab}^z&\equiv& {T\over 2}
{\Bigl[\cosh (z)-\cosh(z\dot G_{ab})\Bigr]\over
z\sinh (z)}
=G_{Bab}-{1\over 3T}G_{Bab}^2z^2+O(z^4)
\label{abbrzGz}\\
\dot G_{Bab}^z&=&{\sinh(z\dot G_{Bab})\over\sinh(z)}
= \dot G_{Bab} -{2\over 3T}\dot G_{Bab}G_{Bab}z^2
+O(z^4)
\label{abbrdotGz}
\ear\no
(compare eq.~(\ref{defGF})).

With these definitions, 
we can then rewrite the various traces
and determinants 
appearing in eqs.~(\ref{Gamma2scal}),
(\ref{Gamma2scalpI})
as

\begin{eqnarray}
{\rm det}^{-{1\over 2}}
\biggl[{\sin(eFT)\over {eFT}}{\bigl(\bar T 
-{1\over 2}
{\cal C}_{ab}
\bigr )}
\biggr]
&=&
{z\over\sinh(z)}
\gamma^{{D\over 2} -1}\gamma^z\nonumber\\
{\rm tr}\Bigl[\ddot{\cal G}_{Bab}\Bigr]
&=&
2D\delta(\tau_a-\tau_b)
-2(D-2){1\over T}-{4\over T}
{z\cosh(z\dot G_{Bab})\over\sinh(z)}
\nonumber\\
{1\over 2}{\rm tr}\dot{\cal G}_{Bab}
{\rm tr}\biggl[
{\dot{\cal G}_{Bab}\over
{\bar T -{1\over 2}{\cal C}_{ab}}}
\biggr]
&=& \half
\biggl[
(D-2)\dot G_{Bab}+2\dot G_{Bab}^z
\biggr]
\biggl[
(D-2)\dot G_{Bab}\gamma
+2\dot G_{Bab}^z
\gamma^z
\biggr]
\nonumber\\
{1\over 2}{\rm tr}
\biggl[
{(\dot {\cal G}_{aa}-\dot {\cal G}_{ab})
(\dot {\cal G}_{ab}-\dot {\cal G}_{bb})
\over
{\bar T -{1\over 2}{\cal C}_{ab}}}
\biggr]
&=&
-\half (D-2)\dot G_{Bab}^2\gamma
-\biggl[
{\dot G^{z2}_{Bab}}
+{4\over T^2}z^2 G_{Bab}^{z2}
\biggr]
\gamma^z
\, .
\label{simplifyremainder}
\end{eqnarray}
\noindent
The term involving
$\delta(\tau_a-\tau_b)$, stemming from $\ddot{\cal G}_{Bab}$, 
can be omitted, since it will not
contribute in dimensional regularization (it corresponds to a
massless tadpole insertion in field theory).

Inserting these expressions into either
eq.~(\ref{Gamma2scal}) or eq.~(\ref{Gamma2scalpI}),
one finds that the resulting integrals suffer from two
kinds of divergences:

\begin{enumerate}

\item
An overall divergence of
the scalar proper-time integral $\int_0^{\infty}dT$
at the lower integration limit.

\item
Divergences of $\int_0^T\, d\tau_a\int_0^T\,d\tau_b$ 
at the point $\tau_a=\tau_b$ where the
photon end points become coincident.

\end{enumerate}
\noindent
The first one must be removed by 
one- and two-loop photon wave function
renormalization, the second one by the one-loop 
renormalization of the scalar mass.

It turns out that this program is easier to carry
out on a certain linear combination of
eqs.~(\ref{Gamma2scal}) and (\ref{Gamma2scalpI}),
namely

\begin{equation}
{\cal L}^{(2)}_{\rm scal}[B]=
{{D-1}\over D}\times
{\rm eq.}
(\ref{Gamma2scal})
+{1\over D}\times
{\rm eq.}
(\ref{Gamma2scalpI})
\label{Gamma2scaloptim}
\end{equation}
\noindent
(A similar simplification can be achieved by
taking the photon insertion in Landau gauge,
though the resulting parameter integrals are not
identical).
Moreover, we rescale to the unit circle,
$\tau_{a,b}=Tu_{a,b}, \bar T = T\hat T$,
and use translation invariance in $\tau$ to
set $\tau_b=0$. Thus in the following
we have
$G_{Bab}=u_a(1-u_a),\dot G_{Bab}=1-2u_a$.

The resulting integral we write

\begin{equation}
{\cal L}^{(2)}_{\rm scal}
[B]=-
{(4\pi )}^{-D}
{e^2\over 2}
\int_0^{\infty}{dT\over T}e^{-m^2T}T^{2-D} 
\int_0^{\infty}d\hat T 
\int_0^1 d u_a
\,
I(z,u_a,\hat T,D)\nonumber\\
\nonumber\\
\label{scaloptimint}
\end{equation}
\no
where the rescaled integrand $I(z,u_a,\hat T,D)$ depends on
$T$ only through $z$.

In contrast to the calculation in proper-time regularization, 
the
$\hat T$ -- integration is nontrivial in dimensional regularization.
It will therefore be easier to extract all subdivergences {\sl before}
performing this integral.
An analysis of the divergence structure shows that the integrand
can be rewritten in the following way,

\begin{equation}
K(z,u_a,D)\equiv
\int_0^{\infty}
d\hat T\,
I(z,u_a,\hat T,D)
=
K_{02}(z,u_a,D)
+f(z,D)G_{Bab}^{1-{D\over 2}}
+
O\bigl(z^4,G_{Bab}^{2-{D\over 2}}\bigr)
\nonumber\\
\label{scaldivextr}
\end{equation}
\no
with
\bear
K_{02}(z,u_a,D)\!\!\!\!&=&
\!\!\!\!
-4{{D-1}\over{D-2}}G_{Bab}^{1-{D\over 2}}
+{2\over 3D(D-2)}
\biggl[
(D-1)(D-4)
G_{Bab}^{1-{D\over 2}}
+(
-2D^2+18D-4
)
G_{Bab}^{2-{D\over 2}}
\biggr]
z^2\nonumber\\
f(z,D)&=&
\!\!\!\!
{D-1\over D(D-2)}
\biggl\lbrack
4D-{2\over 3}
(D-4)z^2
+(8-4D)
{z\over \sinh(z)}
-8
{z^2\cosh(z)\over \sinh^2(z)}
\biggr\rbrack
=O(z^4)\,.
\label{scaldivterms}
\end{eqnarray}
\noindent
$K_{02}$ consists of the terms constant and quadratic
in $z$, which are the only ones causing a divergence at
$T=0$. The
second term is
$O(z^4)$,
so that its integral already converges at $T=0$,
however it diverges at $u_a=0,1$. 

After splitting off these two terms,
the
integral over the remainder is already finite,
so that one can set $D=4$ in its explicit computation. 
For $D=4$ the
$\hat T$ - integral becomes elementary, and yields

\begin{eqnarray}
K(z,u_a,4)&=&
{z\over\sinh(z)}
\Biggl\lbrace
A_0
{\ln ({G_{Bab}/ G_{Bab}^z})
\over{(G_{Bab}-G_{Bab}^z)}}
+
A_1
{\ln ({G_{Bab}/ G_{Bab}^z})
\over{(G_{Bab}-G_{Bab}^z)}^2}
\nonumber\\
&&
+{A_2\over
(G^z_{Bab})(G_{Bab}-G^z_{Bab})}
+{A_3\over
(G_{Bab})(G_{Bab}-G^z_{Bab})}
\Biggr\rbrace\quad ,\nonumber\\
A_0&=&
3\biggl[
2z^2G^z_{Bab}-{z\over\tanh(z)}-1
\biggr]
 \nonumber\\
A_1 &=&
4z^2G_{Bab}^{z2}
+{1\over 2}
(\dot G^{z2}_{Bab}-\dot G_{ab}^2)
 \nonumber\\
A_2&=&
-4z^2G_{Bab}^{z2}
+{1\over 2}
\dot G_{Bab}^z
(\dot G_{Bab}
- \dot G_{Bab}^z)
\nonumber\\
A_3&=&
{1\over 2}
\dot G_{Bab}
(\dot G_{Bab}-
\Gzp)
\; . 
\label{A1A2A3}
\end{eqnarray}
\noindent
The divergences will now be removed by mass and
photon wave function renormalization,

\bear
m^2&=&m_0^2+\delta m_0^2\nonumber\\
e&=&e_0Z_3^{\half}\nonumber\\
B&=&B_0Z_3^{-\half}.
\label{renormalization}
\ear
\no
So far we have worked in the bare regularized theory,
so that
all our previous formulas should,
for the following, be considered
written in terms of $m_0,e_0,B_0$ instead of
$m,e,B$ (note that this leaves $z$ unaffected).

Since we aim at a direct
comparison with previous calculations, 
the renormalization will be done using
on-shell rather than minimal subtraction.
In on-shell subtraction, 
the photon wave function renormalization has 
the effect of simply
removing the $z^2$ - part of $K_{02}$, and the
remaining $z$ - independent term can, of course,
be also discarded.

The removal of the 
divergence caused by the
second term in 
eq.~(\ref{scaldivextr}) 
takes more effort,
and the mere possibility requires a little
conspiration. 

Let us denote the corresponding contribution to
the effective Lagrangian by $G_{\rm scal}(z,D)$. 
For this term the $u_a$ -
integration factors out, yielding

\begin{equation}
\int_0^1\,du_a\,
G_{Bab}^{1-\Dhalf}
=
B\Bigl(2-\Dhalf,2-\Dhalf\Bigr)
=-{4\over\epsilon}+0+O(\epsilon )
\label{uaint}
\end{equation}
\no
where $B$ denotes the Euler Beta-function.

To proceed further, it is essential to note that 
the function
$f(z,D)$ 
can be related to the 
integrand of the scalar one-loop Euler-Heisenberg
Lagrangian, eq.~(\ref{eulheiscal}).
In dimensional regularization, and with
the
two 
terms lowest order in $z$
subtracted out via one-loop photon
wave function renormalization,
this Lagrangian reads

\be
\bar{\cal L}^{(1)}_{\rm scal}[B_0]=
\int_0^{\infty}
{dT\over T}
\e^{-m_0^2T} 
{(4\pi T)}^{-{D\over 2}}
\biggl[
{z\over\sinh(z)}
+{z^2\over 6}-1
\biggr]
\, .
\label{scaloneloopeulhei}
\ee\no
On the other hand, we can rewrite

\bear
f(z,D)&=&
8{D-1\over D(D-2)}
T^{{D\over 2}+1}{d\over dT}
\biggl\lbrace
T^{-{D\over 2}}
\biggl[
{z\over\sinh(z)}+{z^2\over 6}-1
\biggr]
\biggr\rbrace
\, .
\label{rewritef}
\ear
\no
By a partial integration over $T$, we can therefore
reexpress

\bear
\int_0^{\infty}
{dT\over T}{\rm e}^{-m_0^2T}
T^{2-D}
f(z,D)
&=&
8{D-1\over{D(D-2)}}
\biggl\lbrace
m_0^2
\int_0^{\infty}
{dT\over T}{\rm e}^{-m_0^2T}
T^{3-D}
\biggl[
{z\over\sinh(z)}
+{z^2\over 6}-1
\biggr]\nonumber\\
&+&
{D-4\over 2}
\int_0^{\infty}
{dT\over T}{\rm e}^{-m_0^2T}
T^{2-D}
\biggl[
{z\over\sinh(z)}
+{z^2\over 6}-1
\biggr]
\biggr\rbrace
\label{partint}
\ear
\no
(there are no boundary terms since $f(z)=O(z^4)$).

At the two-loop level, the 
effect of mass renormalization
consists in  the following shift
produced by the one-loop 
mass displacement $\delta m_0^2$,

\be
\delta{\cal L}_{\rm scal}^{(2)}
[B_0]
=
\delta
m_0^2
{\partial\over\partial m_0^2}
\bar{\cal L}^{(1)}_{\rm scal}
[B_0]
\, .
\label{scaldeltaL}
\ee\no
$\delta m_0^2$ 
is generated by the UV divergence of the
one-loop scalar self energy in scalar QED.
This quantity
we have to take from
standard field theory. In dimensional
regularization one has
\footnote
{Note that this differs by a sign from $\delta m^2$
as used in ~\cite{ss3}. Here this denotes the
mass displacement itself, there the corresponding counterterm.}

\be
\delta m_0^2=
m_0^2
{\alpha_0\over 4\pi}
\Bigl[
-{6\over\epsilon}
+7
-3
\bigl[\gamma - \ln (4\pi)\bigr]
-3\ln (m_0^2)
\Bigr]+O(\epsilon)
\, .
\label{deltamscal}
\ee
\no
Here $\epsilon = D-4$,
and $\gamma$ denotes the Euler-Mascheroni
constant
\footnote{In comparing with \cite{ritusspin,ditreu,rescsc,ritusscal}
note that there this constant had been denoted by $\ln (\gamma)$.}. 
Expanding eqs.~(\ref{uaint}),
(\ref{partint}), and 
(\ref{scaldeltaL})
in $\epsilon$ one finds that,
up to terms of order $O(\epsilon)$,

\bear
G_{\rm scal}(z,D)&=&
\delta m_0^2 
{\partial\over\partial m_0^2}
\bar{\cal L}^{(1)}_{\rm scal}[B_0]
+m_0^2
{\alpha_0\over{(4\pi)}^3}
\int_0^{\infty}
{dT\over T^2}
\e^{-m_0^2T}
\biggl[
{z\over\sinh(z)}
+{z^2\over 6}-1
\biggr]
\nonumber\\
&&
\times
\biggl[
-3\gamma -3\ln (m_0^2T)+{3\over m_0^2T}
+{9\over 2}
\biggr]\,.
\label{calcG}
\ear
\no
Note that the whole divergence of
$G_{\rm scal}(z,D)$ for $D\rightarrow 4$
has now been absorbed into $\delta m_0^2$.

Our final answer for the
two-loop contribution to the
finite renormalized scalar QED
Euler-Heisenberg thus becomes
\bear
{\cal L}_{\rm scal}^{(2)}[B]
&=&
-{\alpha\over
2{(4\pi )}^{3}}
\int_{0}^{\infty}{dT\over T^3}e^{-m^2T}
\int_0^1 du_a
\,
\Bigl[
K(z,u_a,4)-K_{02}(z,u_a,4)-{f(z,4)\over G_{Bab}}
\Bigr]
\nonumber\\
&&
+{\alpha\over {(4\pi)}^3}m^2
\int_0^{\infty}
{dT\over T^2}{\rm e}^{-m^2T}
\biggl[
{z\over\sinh(z)}
+{z^2\over 6}-1
\biggr]
\biggl[
-3\gamma
-3\ln ( m^2T)+{3\over m^2T}
+{9\over 2}
\biggr]
.
\nonumber\\
\label{Gammascalrenorm} 
\ear
As far as is known to the present authors, the only previous
calculation of the two-loop Euler-Heisenberg for scalar
QED is the one in ~\cite{ritusscal,lebedev}. The parameter
integral given there is rather different from ours, and we have not
succeeded in directly identifying both representations.
However, we have used MAPLE to
expand both formulas in a Taylor expansion in
$B$ up
to order $O(B^{20})$,
and found exact agreement for the coefficients.

\no
Let us just give the first few terms in this 
expansion,

\be
{\cal L}_{\rm scal}^{(2)}[B]
=
{\alpha m^4\over
{(4\pi )}^{3}}
{1\over 81}
\Biggl[
{\displaystyle \frac {275}{8}}
{\Bigl({B\over B_{\rm cr}}\Bigr)}^4
-
{\displaystyle \frac {5159}{200}}
{\Bigl({B\over B_{\rm cr}}\Bigr)}^6
+
{\displaystyle \frac {2255019}{39200}}
{\Bigl({B\over B_{\rm cr}}\Bigr)}^8
-
{\displaystyle \frac {931061}{3600}}
{\Bigl({B\over B_{\rm cr}}\Bigr)}^{10}
+\ldots \,
\Biggr]
.
\label{scalexpand}
\ee


\no
The expansion parameter has been rewritten 
in terms of
$B_{\rm cr}\equiv {m^2\over e}\approx 4.4 \cdot 10^{13}G$.

The corresponding
calculation for spinor QED is completely
analogous.
In the worldline superfield formalism
of ~\cite{rescsc,ss3,polbook}, the
formulas (\ref{Gamma2scal}), (\ref{wickscal})
generalize to the following
integral representation
for the two-loop effective action
induced by the spinor loop,

\begin{eqnarray}
{\cal L}^{(2)}_{\rm spin}
[F]&=&
(-2)
{(4\pi )}^{-D}
\Bigl(-{e^2\over 2}\Bigr)
\int_0^{\infty}{dT\over T}e^{-m^2T}T^{-{D\over 2}} 
\int_0^{\infty}d\bar T 
\int_0^T d\tau_a d\tau_b
\int d\theta_a d\theta_b
\nonumber\\
&\phantom{=}&\times
{\rm det}^{-{1\over 2}}
\biggl[{\tan(eFT)\over {eFT}}
\biggr]
{\rm det}^{-{1\over 2}}
\biggl[
\bar T 
-{1\over 2}
\hat{\cal C}_{ab}
\biggr]
\langle
-D_ay_a\cdot D_by_b\rangle
\quad ,\nonumber\\
\langle
-D_ay_a\cdot D_b y_b\rangle
&=&
{\rm tr}
\biggl[
D_aD_b\hat{\cal G}_{ab}+{1\over 2}
{D_a( \hat{\cal G}_{aa}- \hat{\cal G}_{ab})
D_b( \hat{\cal G}_{ab}- \hat{\cal G}_{bb})
\over
{\bar T -{1\over 2}\hat{\cal C}_{ab}}}
\biggr]\,.
\label{Gamma2spin}
\end{eqnarray}
The $\hat{\cal G}$ appearing here is the 
constant field worldline
superpropagator,

\be
\hat {\cal G}(\tau_1,\theta_1;\tau_2,\theta_2)
\equiv {\cal G}_B(\tau_1,\tau_2) +
\theta_1\theta_2 {\cal G}_F(\tau_1,\tau_2)
\,
\label{superprop}
\ee
\no
which besides the bosonic propagator
eq.~(\ref{defGF}) also contains a fermionic piece,

\be
{\cal G}_{F}(\tau_1,\tau_2) =
{\rm sign} (\tau_1-\tau_2)
{{\rm e}^{-ieFT\dot G_{B12}}\over {\rm cos}(eFT)}
\, .
\label{defGFF}
\ee\no
Our superfield conventions are
$D = {\partial\over{\partial\theta}} - 
   \theta
{\partial\over{\partial\tau}}$,
$\int d\theta\theta = 1$.

Performing the Grassmann integrations, and
removing $\ddot {\cal G}_{Bab}$ by partial integration
over $\tau_a$, we obtain the equivalent of 
eq.~(\ref{Gamma2scalpI}),

\begin{eqnarray}
{\cal L}^{(2)}_{\rm spin}[F]&=&
{(4\pi )}^{-D}
e^2
\int_0^{\infty}{dT\over T}e^{-m^2T}T^{-{D\over 2}} 
\int_0^{\infty}d\bar T 
\int_0^T d\tau_a
\int_0^T d\tau_b
\nonumber\\
&\phantom{=}&\times
{\rm det}^{-{1\over 2}}
\biggl\lbrack{\tan(eFT)\over {eFT}}
\Bigl(
\bar T -{1\over 2}
{\cal C}_{ab}
\Bigr)
\biggr\rbrack
{1\over 2}
\Biggl\lbrace
{\rm tr}\dot{\cal G}_{Bab}
{\rm tr}
\biggl\lbrack
{\dot{\cal G}_{Bab}
\over
{\bar T -{1\over 2}{\cal C}_{ab}}}
\biggr\rbrack
-{\rm tr}{\cal G}_{Fab}
{\rm tr}\biggl[
{{\cal G}_{Fab}\over
{\bar T -{1\over 2}{\cal C}_{ab}}}
\biggr]
\nonumber\\
&\phantom{=}&
+{\rm tr}
\biggl[
{(\dot {\cal G}_{Baa}-\dot {\cal G}_{Bab})
(\dot {\cal G}_{Bab}-\dot {\cal G}_{Bbb}+2{\cal G}_{Faa})
+{\cal G}_{Fab}{\cal G}_{Fab}
-{\cal G}_{Faa}{\cal G}_{Fbb}
\over
{\bar T -{1\over 2}{\cal C}_{ab}}}
\biggr]
\Biggr\rbrace\; .
\label{Gamma2spinpI}
\end{eqnarray}
\noindent
Note that this formula reduces to 
eq.~(\ref{Gamma2scalpI}), if one 
replaces $\tan(eFT)$ by $\sin(eFT)$,
and deletes
all the ${\cal G}_F$, 
as well as the global
factor of $-2$ (which accounts for
the
difference in 
statistics
and degrees of freedom 
between the spin $0$ and
spin $1\over 2$ loops).

Contrary to the scalar QED case, 
here the partially integrated integral is
already a suitable starting point for renormalization
(for more on this point see chapter 7 of
~\cite{rescsc}).

Specializing to the magnetic field case,
it is again easy to
calculate the Lorentz determinants and traces. 
After rescaling to the unit circle, one
obtains a parameter integral

\begin{equation}
{\cal L}^{(2)}_{\rm spin}
[B]=
{(4\pi )}^{-D}
e^2
\int_0^{\infty}{dT\over T}e^{-m^2T}T^{2-D} 
\int_0^{\infty}d\hat T 
\int_0^1 d u_a
\,
J(z,u_a,\hat T,D)
\, .
\nonumber\\
\label{spinoptimint}
\end{equation}
\no
The extraction of the subdivergences
yields

\begin{equation}
L(z,u_a,D)\equiv
\int_0^{\infty}
d\hat T\,
J(z,u_a,\hat T,D)
=
L_{02}(z,u_a,D)
+g(z,D)G_{Bab}^{1-{D\over 2}}
+
O\bigl(z^4,G_{Bab}^{2-{D\over 2}}\bigr)
\nonumber\\
\label{spindivextr}
\end{equation}
\no
with
\bear
L_{02}(z,u_a,D)&=&
-4(D-1)G_{Bab}^{1-{D\over 2}}
-{4\over 3D}
\biggl[
(D-1)(D-4)
G_{Bab}^{1-{D\over 2}}
+(D-2)(D-7)
G_{Bab}^{2-{D\over 2}}
\biggr]
z^2\nonumber\\
g(z,D)&=&
-{4\over 3}
{D-1\over D}
\biggl\lbrack
6{z^2\over\sinh^2(z)}
+3(D-2)z\coth(z)-(D-4)z^2
-3D
\biggr\rbrack
=O(z^4)
\, .
\nonumber\\
\label{spindivterms}
\end{eqnarray}
\noindent
$L_{02}$ is again removed by photon wave function renormalization.
Denoting the contribution of the second term by
$G_{\rm spin}(z,D)$, we note
that the $u_a$ - integral is the same as in the
scalar QED case, eq.~(\ref{uaint}).
Using the following identity 
analogous to eq.~(\ref{rewritef}),

\bear
g(z,D)&=&
8{D-1\over D}
T^{{D\over 2}+1}{d\over dT}
\biggl\lbrace
T^{-{D\over 2}}
\biggl[
{z\over\tanh(z)}-{z^2\over 3}-1
\biggr]
\biggr\rbrace
\label{rewriteg}
\ear
\no
we partially integrate the remaining integral over $T$. 
The $1\over\epsilon$ - part of $G_{\rm spin}$ is then
again found to be just right for absorbing the
shift induced by the one-loop mass displacement,

\be
\delta m_0=
m_0
{\alpha_0\over 4\pi}
\Bigl[
-{6\over\epsilon}
+4
-3
\bigl[\gamma - \ln (4\pi)\bigr]
-3\ln (m_0^2)
\Bigr]+O(\epsilon)
\, .
\label{deltamspin}
\ee
\no
Up to terms of order $\epsilon$ one obtains

\bear
G_{\rm spin}(z,D)&=&
\delta m_0 
{\partial\over\partial m_0}
\bar{\cal L}^{(1)}_{\rm spin}[B_0]
+m_0^2
{\alpha_0\over{(4\pi)}^3}
\int_0^{\infty}
{dT\over T^2}
\e^{-m_0^2T}
\biggl[
{z\over\tanh(z)}
-{z^2\over 3}-1
\biggr]
\nonumber\\
&&
\times
\biggl[
12\gamma +12\ln ( m_0^2T)-{12\over m_0^2T}
-18
\biggr]\,.
\label{calcGspin}
\ear
\no
Our final result for the on-shell renormalized
two-loop spinor QED
Euler-Heisenberg Lagrangian is

\bear
{\cal L}_{\rm spin}^{(2)}[B]
&=&
{\alpha\over
{(4\pi )}^{3}}
\int_{0}^{\infty}{dT\over T^3}e^{-m^2T}
\int_0^1 du_a
\,
\Bigl[
L(z,u_a,4)-L_{02}(z,u_a,4)-{g(z,4)\over G_{Bab}}
\Bigr]\nonumber\\
&&-
{\alpha\over {(4\pi)}^3}
m^2\int_0^{\infty}
{dT\over T^2}{\rm e}^{-m^2T}
\biggl[
{z\over\tanh(z)}
-{z^2\over 3}-1
\biggr]
\biggl[18-12\gamma-
12\ln (m^2T)+{12\over m^2T}
\biggr]
\nonumber\\
\label{Gammaspinrenorm}
\ear\no
with

\bear
L(z,u_a,4)&=&{z\over\tanh(z)}
\Biggl\lbrace
B_1
{\ln ({G_{Bab}/G_{Bab}^z})
\over{(G_{Bab}-G_{Bab}^z)}^2}
+{B_2\over
G^z_{Bab}(G_{Bab}-G^z_{Bab})}
+{B_3\over
G_{Bab}(G_{Bab}-G^z_{Bab})}
\Biggr\rbrace\nonumber\\
B_1&=&4z
\Bigl(
\coth(z)-\tanh(z)
\Bigr)
G^z_{Bab}-4G_{Bab}
\quad  \nonumber\\
B_2&=&2\dot G_{Bab}\dot G^z_{Bab}+ 
z(8\tanh(z)-4\coth(z))
G^z_{Bab}-2
\quad \nonumber\\
B_3&=&4G_{Bab}
-2\dot G_{Bab}\dot G^z_{Bab}
-4z\tanh(z)G^z_{Bab}+2\nonumber\\
L_{02}(z,u_a,4)&=&-{12\over G_{Bab}}+2z^2\nonumber\\
g(z,4)&=&-6\biggl[
{z^2\over{\sinh(z)}^2}+z\coth(z)-2
\biggr].
\label{defLLg}
\ear\no
Comparing with the previous results by Ritus 
and Dittrich-Reuter, we have again not succeeded
in a direct identification with the 
more complicated parameter integral given
by Ritus \cite{ritusspin}.
However, as in the scalar QED case we have verified agreement
between both formulas up to the
order of $O(B^{20})$ in the weak-field
expansion in $B$. The first few coefficients are

\be
{\cal L}_{\rm spin}^{(2)}[B]
=
{\alpha m^4\over
{(4\pi )}^{3}}
{1\over 81}
\Biggl[
64 
{\Bigl({B\over B_{\rm cr}}\Bigr)}^4
-{1219\over 25}
{\Bigl({B\over B_{\rm cr}}\Bigr)}^6
+ {135308\over 1225}
{\Bigl({B\over B_{\rm cr}}\Bigr)}^8
-{791384\over 1575}
{\Bigl({B\over B_{\rm cr}}\Bigr)}^{10}
+\ldots \, 
\Biggr]
.
\label{spinexpand}
\ee


On the other hand, our formula {\sl almost}
allows for a term by term
identification with the result of Dittrich-Reuter
\cite{ditreu}, as given in eqs.~(7.21),(7.22) there.
This requires 
a rotation to Minkowskian proper-time,
$T\rightarrow is$, a transformation of variables
from $u_a$ to $v:=\dot G_{Bab}$, 
the use of trigonometric identities,
and another partial
integration over $T$ for the last two terms in
eq.~(\ref{Gammaspinrenorm}). The only discrepancy arises in
the constant $18$, which reads $10$ in the Dittrich-Reuter
formula. 

Since this constant can be adjusted by a change of the
finite constant appearing in $\delta m_0$,
we conclude that the two previous results for this
effective Lagrangian differ precisely by a finite mass
renormalization. Moreover, it is clear 
that Ritus' formula is the one which 
correctly identifies the physical electron mass.

Finally, let us mention that
the one-loop
Euler-Heisenberg Lagrangian, 
and perhaps even its two-loop
correction considered here, may possibly be
measured in optical experiments in the near
future \cite{bakalovetal}.

\vskip15pt
\noindent
{\bf Acknowledgements:}
C.S. would like to thank O. Tarasov for discussions, and
A.A. Vladimirov for correspondence.

\vfill\eject

\vfill\eject


\begin{thebibliography}{99}
\bibitem{eulhei}W. Heisenberg, H. Euler, Z. Phys. {\bf 38} 
(1936) 714.
\bibitem{schwinger51}
J. Schwinger, Phys. Rev. {\bf 82} (1951) 664.
\bibitem{ritusspin}
V. I. Ritus, Sov. Phys. JETP {\bf 42} (1975) 774.
\bibitem{ditreu}W. Dittrich and M. Reuter, {\it
Effective Lagrangians in Quantum Electrodynamics},
Springer 1985.
\bibitem{rescsc}M. Reuter, M.G. Schmidt and C. Schubert,
IASSNS-HEP-96/90 (hep-th/9610191), {\sl Ann. Phys.}
(in press).
\bibitem{strassler}M. J. Strassler, Nucl. Phys. {\bf B385} 
(1992) 145.
\bibitem{ss1}M. G. Schmidt, C. Schubert, Phys. 
Lett. {\bf B318} (1993) 438 (hep-th/9309055).
\bibitem{ss2}M. G. Schmidt, C. Schubert, Phys. 
Lett. {\bf B331} (1994) 69 (hep-th/9403158).
\bibitem{ss3}M. G. Schmidt, C. Schubert, Phys. Rev. {\bf D53}
(1996) 2150 (hep-th/9410100).
\bibitem{dashsu}K. Daikouji, M. Shino, Y. Sumino, 
Phys. Rev. {\bf D53} (1996) 4598
(hep-ph/9508377). 
\bibitem{rolsat}K. Roland, H. Sato, Nucl. Phys. {\bf B 480}
(1996) 99 
(hep-th/9604152).   
\bibitem{zako}C. Schubert, Lectures given
at the XXXVI Cracow School of Theoretical Physics,
Act. Phys. Pol. {\bf B 27} (1996) 3965
(hep-th/9610108).
\bibitem{bk}Z. Bern, D. A. Kosower,
Phys. Rev. Lett. {\bf 66} (1991) 1669; 
\hfill\break
Nucl. Phys. {\bf B379} (1992)
451.
\bibitem{berdun}
Z. Bern, D. C. Dunbar, Nucl. 
Phys. {\bf B379} (1992)
562.
\bibitem{vladimirov}A.A. Vladimirov, Teor. Mat. Fiz. {\bf 36}
(1978) 271.
\bibitem{feynman}R. P. Feynman, Phys. Rev. {\bf 80} (1950)
  440.
\bibitem{adlsch}S. L. Adler and C. Schubert, Phys. Rev. Lett.
{\bf 77} (1996) 1695
(hep-th/9605035).
\bibitem{cdd} D. Cangemi, E. D'Hoker, G. Dunne, 
Phys. Rev. {\bf D51} (1995) 2513 (hep-th/9409113).
\bibitem{shaisultanov}
R. Shaisultanov, Phys. Lett. {\bf B 378} (1996) 354 (hep-th/9512142).
\bibitem{ritusscal}V. I. Ritus, Zh. Eksp. Teor. Fiz. {\bf 73}
(1977) 807 [Sov. Phys. JETP {\bf 46} (1977) 423].
\bibitem{lebedev}S. L. Lebedev, FIAN No. 254, Moscow 1982 (unpublished).
\bibitem{polbook}A. M. Polyakov, {\sl Gauge Fields and Strings},
   Harwood 1987.
\bibitem{bakalovetal} D. Bakalov et al., Nucl. Phys. {\bf B35}
(Proc. Suppl.) (1994) 180;\hfill\break
D. Bakalov, INFN/AE-94/27 (unpublished).

\end{thebibliography}
\end{document}